\documentclass[12pt]{article}
\usepackage{a4}
\usepackage{amsfonts}
\usepackage{amsmath}
\usepackage{latexsym}
\usepackage{amsfonts}
\usepackage{graphics}
\usepackage{graphicx}
\usepackage{epic}
\usepackage{eepic}

\begin{document}

\noindent arXiv:1002.1358 [hep-th]
   \hfill  November 2010 \\

\renewcommand{\theequation}{\arabic{section}.\arabic{equation}}
\thispagestyle{empty}
\vspace*{-1,5cm}
\noindent \vskip3.3cm

\begin{center}
{\Large\bf Direct construction of a cubic selfinteraction for higher spin gauge fields}

{\large Ruben Manvelyan ${}^{\dag\ddag}$, Karapet Mkrtchyan${}^{\ddag}$ \\and Werner R\"uhl
${}^{\dag}$}
\medskip

${}^{\dag}${\small\it Department of Physics\\ Erwin Schr\"odinger Stra\ss e \\
Technical University of Kaiserslautern, Postfach 3049}\\
{\small\it 67653
Kaiserslautern, Germany}\\
\medskip
${}^{\ddag}${\small\it Yerevan Physics Institute\\ Alikhanian Br.
Str.
2, 0036 Yerevan, Armenia}\\
\medskip
{\small\tt manvel,ruehl@physik.uni-kl.de; karapet@yerphi.am}
\end{center}\vspace{2cm}

\bigskip
\begin{center}
{\sc Abstract}
\end{center}
\quad  Using Noether's procedure we directly construct a complete cubic selfinteraction for the case of spin $s=4$ in a flat background and discuss the cubic selfinteraction for general spin $s$ with $s$ derivatives in the same background. The leading term of the latter interaction together with the leading gauge transformation of first field order are presented.

\newpage

\section*{Introduction}

\quad Several trilinear interactions of higher spin fields where presented in our previous article \cite{M}. Interactions are constructed in the Lagrangian framework using Noether's procedure together with the corresponding gauge field transformations next to the free level. In this article we turn to the construction of the cubic selfinteraction for the spin $s$ higher spin gauge field in a flat background using again direct Noether's procedure.

The construction of (self)interacting \emph{higher spin gauge field} theories was always in the center of attention during the last thirty years. Without aiming at a complete list of literature we just refer here to those articles that are important for our investigation \cite{vanDam}-\cite{Petkou}, a more complete reference list could be easily obtained from these. The first big step into the theory of linearly gauge invariant cubic interaction Lagrangians ("CILs")of higher spin fields was done in \cite{vanDam}
where the fields have equal spin three, three derivatives are applied, and the nonexistence of gauge transformations of higher than first order was proved. Recent considerations of this spin 3 selfinteraction have appeared in \cite{bekaert}. It was shown that the Berends-Burgers-van Dam vertex does not allow higher order continuations even if we take into account possible interactions with other fields of spin higher (or lower) than three. This is not surprising, because taking into account Metsaev's formula for the possible numbers of derivatives in CILs, one can easily see that \emph{the minimal selfinteraction of HS gauge field of any spin does not include corrections from interactions with fields of another spin value in flat space}, but in constantly curved backgrounds like AdS or dS this (spin 3) selfinteraction have a good chance to be continued to higher orders due to the corrections from interactions with fields of different spin. It is also shown in \cite{bekaert}, that for a configuration ($s_{1},s_{2},s_{3}$) with $s_{3} \leq s_{2} \leq s_{1}$, the cubic vertices containing up to $2s_{2} - 1$ derivatives (resp. $2s_{2}-2$) for an odd sum $s_{1} + s_{2} + s_{3}$ (resp. for an even sum $s_{1} + s_{2} + s_{3}$) give rise to a non-abelian gauge algebra at first nontrivial order in the deformation, while the vertices with more derivatives are abelian at the same order in full agreement with \cite{M}. In other words if the CIL includes derivatives less than $2s_2$, it can't be constructed only from curvatures for $s_2$ and $s_3$, therefore it is nonabelian. Also, a vertex such that $s_{1}\geq s_{2}+s_{3}$ is necessarily abelian at the same first non-trivial order in perturbation, which is obvious because in that case the minimal number of derivatives is bigger or equal to $2s_2$.

Our work presented in this article is a natural continuation of the work \cite{vanDam}. Whereas the authors of \cite{vanDam} do not describe the methods they used to derive the CIL, we develop a recursive formalism, test it on the spin two case (gravity), and then apply it to the spin four selfinteraction successfully. It can of course be used for any triplet of (even) spins.

CILs for a triplet of spins $s_1,s_2,s_3$ depend on a fourth parameter, the number $\Delta$ of derivatives. For each such triplet
there is a minimal number of these, $\Delta_{min}$, which was derived in \cite{Metsaev}.
If $\Delta = s_1+s_2+s_3$ a CIL is obtained by any contraction of the three curvatures of the fields. This is considered as trivial. In all cases known explicitly the CIL is unique due to partial integration and field redefinition. Less trivial are those cases where two fields have equal spin and the respective Bell-Robinson conserved current \cite{vanDam} enters the CIL (see \cite{M}). In another group of cases the Weyl tensor of one of the spins enters the CIL \cite{boulanger}. The CILs for selfinteractions belong to the most complicated category, and the spin four case with $\Delta= \Delta_{min}=4$ treated here is among the simplest of these.

Although it was believed that higher-spin gravitational interaction in flat space is inconsistent \cite{Deser}, the very important considerations for CIL's which describe Higher Spin couplings to gravity and general $s,s^\prime,s^{\prime\prime}\geq 1$ in four dimensions are discussed in \cite{Vasiliev}\footnote{We would like to note that Fradkin-Vasiliev results \cite{Vasiliev} obtained in the \emph{four-dimensional AdS} space overlap only partially with our results obtained \emph{in any dimensional flat space-time} \cite{ourlast}.}. It is argued by the authors that Higher Spin cubic couplings do not have a flat limit. Nevertheless it was shown for spin four coupling to gravity in \cite{boulanger} that after appropriate rescaling of the spin four fields one can get a flat limit of the Fradkin-Vasiliev vertex with six derivatives in full agreement with \cite{Metsaev} and the same was conjectured for any higher spin interaction of Fradkin-Vasiliev (2-s-s) type as well as for nonabelian interaction of type 1-s-s.
The present status of applications of Vasiliev's full nonlinear theory is represented e.g. in \cite{Giombi}. These authors
analyze the equations of motions with respect to their physical field components. But at the end the authors determine only the relative normalization constants $C(s_1,s_2,s_3)$ of the CILs, not the CILs themselves, and this also only for cases when at least one spin is zero. In the future when all CILs will be known explicitly (we believe to be close to this), we may insert these
into the equations of motion as presented e.g. by \cite{Giombi} and thus investigate their solvability.

Historically  this complicated task of general field theory always attracted interest but activity intensified after discovering the important role higher spin gauge fields play in $AdS/CFT$ correspondence especially after discovering the holographic duality between the $O(N)$ sigma model in three dimensional space and higher spin gauge theory operating in the four dimensional space with negative constant curvature \cite{Kleb}. This case of holography is singled out by the existence of two conformal points, notably weakly and strongly coupled, of the
boundary theory and the possibility to describe them by the
same higher spin gauge theory with the help of spontaneously breaking of higher spin gauge symmetry and mass generation by a corresponding Higgs mechanism. These complicated quantum field theory tasks relate to \emph{quantum loop} calculations for higher spin fields  \cite{MMR1}-\cite{MR5} and therefore necessitate the existence of different possible interactions of higher spin fields
that are manifest, off-shell and formulated in a Lagrangian framework. Successful interaction constructions could be applied for example to one loop calculations. On the quantum level this construction can be controlled by comparison with the boundary $O(N)$ model enabling us to check the $AdS/CFT$ correspondence conjecture on the loop level \cite{MMR1}, \cite{Ruehl}, \cite{MR2}. On the other hand one loop calculations are mainly interesting in the framework of their ultraviolet behaviour when the difference between an $AdS$ and a flat space background can be neglected at least in the leading order.

In this article we continue the issue to construct possible couplings  which we started in our previous articles that involved couplings among different higher spin fields and scalar fields \cite{M, MMR, MR, MMR1} . Here we turn to the trilinear or cubic
selfinteraction of  Fronsdal's \cite{Frons} spin $s$ gauge fields in a flat background but the results can in principle be generalized to the $AdS$ background.

The first sections are devoted to the development of the idea: how can we apply Noether's equation to construct a spin $s$ gauge field selfinteraction in an algorismized manner with a useful classification for the parts of the interaction Lagrangian. After the development of the formalism and corresponding technique for solving the functional Noether's equation, we formulate a general prediction for the leading terms of the spin $s$  cubic interaction Lagrangian and the leading part of the gauge transformation next to the free terms (linear in the gauge field). In section two we recall the gravity case as an exercise and present and analyze in detail the complete solution for the $s=2$ cubic interaction having in mind to find a general ansatz for the solution of Noether's equation.
In section three we present our main result of this article: \emph{the complete solution for the cubic selfinteraction in the case of spin four}. The result includes both the cubic Lagrangian and the part of the gauge transformation linear in the field. The leading parts of these formulas are in full agreement with our prediction for general spin $s$ formulated in the first section.

\section{Higher spin gauge field selfinteraction: The beauty of Noether's procedure }
\quad Following our previous articles \cite{MR1}-\cite{MR5} we use  the most elegant and convenient  way of handling symmetric tensors such as $h^{(s)}_{\mu_1\mu_2...\mu_s}(z)$ by contracting it with the $s$'th tensorial power of a vector $a^{\mu}$ of the tangential space at
the base point $z$
\begin{equation}
h^{(s)}(z;a) = \sum_{\mu_{i}}(\prod_{i=1}^{s} a^{\mu_{i}})h^{(s)}_{\mu_1\mu_2...\mu_s}(z) .
\label{2.5}
\end{equation}
In this way for spin $s$ we obtain a homogeneous polynomial in the vector $a^{\mu}$ of degree $s$.
Then we can write the symmetrized gradient, trace and divergence \footnote{To distinguish easily between "a" and "z" spaces we introduce for space-time derivatives $\frac{\partial}{\partial z^{\mu}}$ the notation $\nabla_{\mu}$.}
\begin{eqnarray}
&&Grad:h^{(s)}(z;a)\Rightarrow Gradh^{(s+1)}(z;a) = (a\nabla)h^{(s)}(z;a) , \\
&&Tr:h^{(s)}(z;a)\Rightarrow Trh^{(s-2)}(z;a) = \frac{1}{s(s-1)}\Box_{a}h^{(s)}(z;a) ,\\
&&Div:h^{(s)}(z;a)\Rightarrow Divh^{(s-1)}(z;a) = \frac{1}{s}(\nabla\partial_{a})h^{(s)}(z;a) .
\end{eqnarray}
All other  manipulations in this formalism are discussed in the Appendix A of this paper. Here we will only present
Fronsdal's Lagrangian in terms of these conventions:
\begin{equation}\label{4.42(1)}
 \mathcal{L}_{0}(h^{(s)}(a))=-\frac{1}{2}h^{(s)}(a)*_{a}\mathcal{F}^{(s)}(a)
    +\frac{1}{8s(s-1)}\Box_{a}h^{(s)}(a)*_{a}\Box_{a}\mathcal{F}^{(s)}(a) .
\end{equation}
where $\mathcal{F}^{(s)}(z;a)$ is the so-called Fronsdal tensor
\begin{eqnarray}
\mathcal{F}^{(s)}(z;a)=\Box h^{(s)}(z;a)-s(a\nabla)D^{(s-1)}(z;a) \quad\label{4.32(1)}
\end{eqnarray}
and $D^{(s-1)}(z;a)$ is the so-called de Donder tensor or traceless divergence of the higher spin gauge field
\begin{eqnarray}\label{4.42(D)}
 && D^{(s-1)}(z;a) = Divh^{(s-1)}(z;a)
-\frac{s-1}{2}(a\nabla)Trh^{(s-2)}(z;a)\\
&& \Box_{a} D^{(s-1)}(z;a)=0
\end{eqnarray}
The initial gauge variation of a spin $s$ field that is of field order zero is
\begin{eqnarray}\label{4.5}
\delta_{(0)} h^{(s)}(z;a)=s (a\nabla)\epsilon^{(s-1)}(z;a) ,
\end{eqnarray}
with the traceless gauge parameter
\begin{eqnarray}
\Box_{a}\epsilon^{(s-1)}(z;a)=0 ,\label{4.6}
\end{eqnarray}
for the by definition double traceless gauge field
 \begin{eqnarray}
\Box_{a}^{2}h^{(s)}(z;a)=0 .
\end{eqnarray}
Therefore on this level we can see from (\ref{4.5}) and (\ref{4.6})  that a correct generalization of the Lorentz gauge condition in the case of $s>2$ could be only the so-called de Donder gauge condition
\begin{equation}\label{dd}
    D^{(s-1)}(z;a)=0 .
\end{equation}

The equation of motion following  from (\ref{4.42(1)}) is
\begin{equation}\label{4.45}
     \delta\mathcal{L}_{0}(h^{(s)}(a))=-(\mathcal{F}^{(s)}(a)-\frac{a^{2}}{4}\Box_{a}\mathcal{F}^{(s)}(a))*_{a}\delta h^{(s)}(a) ,
\end{equation}
and zero order gauge invariance (when $\delta h^{(s)}(a)=\delta_{(0)}h^{(s)}(a)$) can be  checked by substitution of (\ref{4.5}) into this variation and use of the duality relation (\ref{4.15}) and identity (\ref{4.42}) taking into account tracelessness  of the gauge parameter (\ref{4.6}).

Now we turn to the formulation of Noether's general procedure for constructing the spin $s$ cubic selfinteraction.
Similar to  \cite{vanDam} Noether's equation in this case looks like\footnote{From now on we will admit integration everywhere where it is necessary (we work with a Lagrangian as with an action) and therefore we will neglect all $d$ dimensional space-time total derivatives when making a partial integration.}
\begin{equation}\label{4.58}
    \delta_{(1)}\mathcal{L}_{0}(h^{(s)}(a))+\delta_{0}\mathcal{L}_{1}(h^{(s)}(a))=0 .
\end{equation}
where $\mathcal{L}_{1}(h^{(s)}(a))$ is a cubic interaction Lagrangian and $\delta_{(1)}h^{(s)}(a)$ is a gauge transformation that is of first order in the gauge field. Actually equation (\ref{4.58}) just expresses in the cubic order of the field the generalized gauge invariance
\begin{equation}\label{gge}
    \delta \mathcal{L}(h^{(s)}(a))=\frac{\delta \mathcal{L}(h^{(s)}(a)}{\delta h^{(s)}(a))}*_{a}\delta h^{(s)}(a)=0
\end{equation}
where
\begin{eqnarray}
  \mathcal{L}(h^{(s)}(a)) &=& \mathcal{L}_{0}(h^{(s)}(a)) + \mathcal{L}_{1}(h^{(s)}(a)) + \dots \\
  \delta h^{(s)}(a)  &=& \delta_{(0)} h^{(s)}(a)+ \delta_{(1)}h^{(s)}(a)+ \dots
\end{eqnarray}
Combining (\ref{4.45}) and (\ref{4.58}) we obtain the following functional Noether's equation
\begin{equation}\label{4.481}
    \delta_{(0)}\mathcal{L}_{1}(h^{(s)}(a))=(\mathcal{F}^{(s)}(a)-\frac{a^{2}}{4}\Box_{a}\mathcal{F}^{(s)}(a))*_{a}\delta_{(1)} h^{(s)}(a)
\end{equation}
and \emph{we would like to present in this article the solution of the latter equation for the case $s=4$ and propose a generalization for any even $s$}.

First we investigate a first order variation of the spin $s$ gauge transformation. Remembering that Fronsdal's higher spin gauge potential  has scaling dimension $\Delta_{s}=s-2$ (zero for the $s=2$ graviton case) and ascribing the same dimensions to the free part of the Lagrangian that is quadratic in the fields and derivatives $\mathcal{L}_{0}(h^{(s)}(a))$ and to the interaction $\mathcal{L}_{1}(h^{(s)}(a))$ cubic in the fields, we arrive at the idea that \emph{the number of derivatives in the interaction should be $s$}. This type of interacting theories will behave in the same way as gravity. Then we can easily conclude from (\ref{4.58}) that the number of derivatives in the first order variation $\delta_{(1)}h^{(s)}(a)$ should be $s-1$. For $s=2$ this consideration is of course in full agrement with the linearized expansion of the Einstein-Hilbert action.

The next observation is connected with double tracelessness of Fronsdal's higher spin gauge potential. This means that we must
make sure that the same holds for the variation. Expanding the general variation in powers of $a^{2}$
\begin{equation}\label{4.46}
    \delta_{(1)} h^{(s)}(a)=\delta_{(1)} \tilde{h}^{(s)}(a)+a^{2}\delta_{(1)} h^{(s-2)}(a)+(a^{2})^{2}\delta h^{(s-4)}(a)+ \dots ,
\end{equation}
we see that the double tracelessness condition $\Box^{2}_{a} \delta h^{(s)}(a)=0$ expresses the third and
higher terms of the expansion (\ref{4.46}) through the first two free parameters $\delta_{(1)} h^{(s)}(a)$ and $\delta_{(1)} h^{(s-2)}(a)$\footnote{For completeness we present here the solution for $\delta h^{(s-4)}(a)$ following from the double tracelessness condition
\begin{eqnarray}
&& \delta h^{(s-4)}(a)=-\frac{1}{8\alpha_{1}\alpha_{2}} \left[\Box^{2}_{a}\delta h^{(s)}_{(1)}(a)+4\alpha_{1}\Box_{a}\delta h^{(s-2)}(a)\right] ,\nonumber\\
  && \alpha_{k}=d+2s-(4+2k) ,\quad k\in \{1,2\} . \nonumber
\end{eqnarray} }.
From the other hand Fronsdal's tensor (and the r.h.s of (\ref{4.481})) is double traceless by definition and therefore all these $O(a^{4})$ terms are unimportant because they do not contribute to (\ref{4.481}). This leaves us freedom in the choice of initial $\delta_{(1)} h^{(s-2)}(a)$. Using this freedom we can shift the initial first order variation in the following way ($d$ denotes the space-time dimension)
\begin{equation}\label{4.47}
    \delta_{(1)} h^{(s)}(a)\Rightarrow \delta_{(1)} h^{(s)}(a)+\frac{a^{2}}{2(d+2s-2)}\Box_{a}\delta h^{(s)}_{(1)}(a) ,
\end{equation}
and discover that (\ref{4.481}) reduces to
\begin{equation}\label{4.482}
     \delta_{(0)}\mathcal{L}_{1}(h^{(s)}(a))=\mathcal{F}^{(s)}(a)*_{a}\delta h^{(s)}_{(1)}(a) .
\end{equation}
Now to solve this equation we can formulate the following strategy:

1) First we can start from any  cubic ansatz with $s$ derivatives $\mathcal{L}_{1}(h^{(s)}(a))$ suitable in respect to the zero order variation (\ref{4.5}) and variate it inserting in the l.h.s. of (\ref{4.482})  .

2) Then we make a partial integration  and rearrange indices to extract an integrable part due to terms proportional to Fronsdal's tensor $\mathcal{F}^{(s)}(a)$ (or $Tr\mathcal{F}^{(s)}(a))$ in agreement with the r.h.s. of (\ref{4.482}).

3) Symmetrizing expressions in this way we classify terms as
\begin{itemize}
  \item integrable
  \item integrable and subjected to field redefinition (proportional to Fronsdal's tensor)
  \item non integrable but reducible by deformation of the initial ansatz for the gauge transformation (again proportional to Fronsdal's tensor)
\end{itemize}

Then if no other terms remain we can construct our interaction together with the corrected first order transformation.
Following this strategy we will consider the $s=2$ and $s=4$ cases in the next sections in detail.  The exact and unique results after field redefinition and partial integration that are presented in the next two sections are in full agreement with the prediction for general even spin $s$.
To formulate this prediction let us first introduce a classification of cubic monoms with $s$ derivatives. We will call leading terms all those monoms without traces and divergences or equivalently without $\bar{h}^{(s-2)}=Tr: h^{(s)}$  and $D^{(s-1)}$, where the derivatives are contracted only with gauge fields and not with other derivatives. This type of terms is interesting because any partial integration will map such term to the terms of the same type and create one additional term with a divergence, which we can map to $D^{(s-1)}$ dependent and trace dependent terms. Another important point of this class of monoms is that inside of this class we have the following important term involving the linearized Freedman-de Witt gauge invariant curvature \cite{DF, MR6}
\begin{eqnarray}
    &&\mathcal{L}^{initial}_{1}(h^{(s)}(a))=\frac{1}{2s}h^{(s)}(b)*_{b}\Gamma^{(s)}(b,a)*_a h^{(s)}(a),\label{impterm}\\
    &&\Gamma^{(s)}(z;b,a)=\sum_{k=0}^{s}\frac{(-1)^{k}}{k!}(b\nabla)^{s-k}(a\nabla)^{k}(b\partial_{a})^{k}h^{(s)}(z;a)\label{curv}
\end{eqnarray}
This term we can use (and we used it in the case s=4) as an initial ansatz for the solution of (\ref{4.482}). Using (\ref{A22}) and (\ref{4.41}) we see that
\begin{equation}\label{variat}
    \delta_{(0)}\mathcal{L}^{initial}_{1}(h^{(s)}(a))=-\epsilon^{(s-1)}(z;b)(b\nabla)h^{(s)}(a)*_a *_b \Gamma^{(s)}(b,a) + O(\mathcal{F}^{(s)})
\end{equation}
It is easy to see from (\ref{curv}) that after variation in the r.h.s. of  (\ref{variat}) we get $s+1$ monoms linear on the gauge parameter $\epsilon^{(s-1)}(z;b)$ and quadratic in the gauge field, where some of them contain two factors $(b\nabla)$ of contracted derivatives. These terms we can separate as next level terms including the de Donder tensor $ D^{(s-1)}(z;b)$.  To prove this statement we note first that due to partial integration there is the following simple formula:
\begin{equation}\label{mumu}
    F(z)\nabla_{\mu}G(z)\nabla^{\mu}H(z)=\frac{1}{2}\left(\Box F(z) G(z) H(z) - F(z)\Box G(z) H(z)- F(z) G(z)\Box H(z)\right)
\end{equation}
The objects $ F(z), G(z), H(z)$ in our case are proportional to $h^{(s)}(z;a)$ or $\epsilon^{(s-1)}(z;a)$. Then using the definition of Fronsdal's operator (\ref{4.32(1)})and from (\ref{4.42(D)}) and (\ref{4.5}) follows the transformation rule
\begin{equation}\label{trr}
    \delta_{(0)}D^{(s-1)}(z;a)=\Box \epsilon^{(s-1)}(z;a)
\end{equation}
This implies that we can classify all terms with contracted derivatives (i.e. terms with Laplacians) as monoms containing $D^{(s-1)}(z;a)$ or $ \delta_{(0)}D^{(s-1)}(z;a)$ which therefore vanish in the de Donder gauge. Actually according to the r.h.s of (\ref{4.482}) we can during functional integration always replace any $\Box h^{(s)}(a)$ with $\mathcal{F}^{(s)}(a)+s(a\nabla)D^{(s-1)}(a)$ obtaining a contribution to $\delta_{(1)}$ and shifting this monom to the next level class comprising one more order of the de Donder tensor.

Operating in this way we can integrate Noether's equation (\ref{4.482}) (or equivalently express the r.h.s. of (\ref{variat}) as   $-\delta_{(0)}\mathcal{L}^{cubic}_{1}(h^{(s)})+O(\mathcal{F}^{(s)})$) using the initial ansatz (\ref{variat}) step by step: integrating first the leading terms without any de Donder tensor or trace, then integrate terms involving only traces but not $D^{(s-1)}(z;a)$. That is the solution in de Donder gauge. After that we can continue the integration and obtain terms linear on $D^{(s-1)}(z;a)$ , quadratic and so on. The procedure will be closed when we obtain a sufficient number of $D^{(s-1)}(z;a)$ to stop the production of terms with contracted derivatives and therefore the production of new level terms coming from formula (\ref{mumu}).

\emph{Collecting the leading terms and rearranging by partial integration derivatives in a cyclic way so that each derivative acting on a tensor gauge field is contracted with the preceding tensor} we finally come to the following prediction for the leading terms of the interaction for a general spin $s$ gauge field:
\begin{eqnarray}
&&\mathcal{L}^{leading}_{(1)}(h^{(s)}(z))=\frac{1}{3s(s!)^{3}}\sum_{\alpha+\beta+\gamma = s}\binom{s}{\alpha,\beta,\gamma}\int_{z_{1},z_{2},z_{3}} \delta(z-z_{1})\delta(z-z_{2})\delta(z-z_{3})\nonumber\\
&&\left[(\nabla_{1}\partial_{c})^{\gamma}(\nabla_{2}\partial_{a})^{\alpha}(\nabla_{3}\partial_{b})^{\beta}
(\partial_{a}\partial_{b})^{\gamma}(\partial_{b}\partial_{c})^{\alpha}(\partial_{c}\partial_{a})^{\beta}\right]h(a;z_{1})h(b;z_{2})h(c;z_{3})\label{prediction}
\end{eqnarray}
where the relative coefficients between monoms are trinomial coefficients:
\begin{equation}\label{trinom}
    \binom{s}{\alpha,\beta,\gamma}=\frac{s!}{\alpha!\beta!\gamma!}, \quad s=\alpha+\beta+\gamma
\end{equation}
Correspondingly the leading term of the first order gauge transformation should be
\begin{eqnarray}
&&\delta_{(1)}^{leading}h^{(s)}(c;z)=\frac{1}{s!(s-1)!}\sum_{\alpha+\beta+\gamma = s}(-1)^{\beta}\binom{s-1}{\alpha-1,\beta,\gamma}\int_{z_{1},z_{2}} \delta(z-z_{1})\delta(z-z_{2})\nonumber\\
&&\ \ \ \ \ \ \left[(c\nabla_{1})^{\gamma}(\nabla_{2}\partial_{a})^{\alpha-1}(\nabla_{1}\partial_{b})^{\beta}
(\partial_{a}\partial_{b})^{\gamma}(c\partial_{b})^{\alpha}(c\partial_{a})^{\beta}\right]\epsilon(a;z_{1})h(b;z_{2})
\end{eqnarray}
Splitting the trinomial into  two binomials we can rewrite this expression in a more elegant way
\begin{eqnarray}
  &&\delta_{(1)}^{leading}h^{(s)}(c;z)=\frac{1}{s!}\sum_{k=0}^{s-1}k!\binom{s-1}{k}\gamma^{(k)}_{(\epsilon^{(s-1)})}(c,b;a)*_{a,b}
(a\nabla)^{s-k-1}(c\partial_{b})^{s-k}h^{(s)}(b)\nonumber\\\label{pred1}
\end{eqnarray}
where
\begin{eqnarray}
&&\gamma^{(k)}_{(\epsilon^{(s-1)})}(c,b;a)\nonumber\\
&&\ \ \ \ =
    \frac{k!}{(s-1)!}\sum_{i=0}^{k}\frac{(-1)^{i}}{i!}(c\nabla)^{k-i}(b\nabla)^{i}(c\partial_{b})^{i}
    \left[(a\partial_{b})^{s-1-k}\epsilon^{(s-1)}(b)\right]\label{gamma}
\end{eqnarray}
Comparring with (\ref{curv}) we see that
\begin{equation}\label{gamma1}
    \gamma^{(k)}_{(\epsilon^{(s-1)})}(c,b;a)=\Gamma^{(k)}(b,c;h^{(k)}_{a}(c)) ,
\end{equation}
where
\begin{equation}\label{gamma2}
 h^{(k)}_{a}(c)=\frac{k!}{(s-1)!}\left[(a\partial_{b})^{s-1-k}\epsilon^{(s-1)}(b)\right] ,
\end{equation}
and therefore the $ \gamma^{(k)}_{(\epsilon^{(s-1)})}(c,b;a)$ coefficients inherit in the $c,b$ index spaces all properties of the corresponding spin $k$ curvature described in details in Appendix A.
In the next two sections we show that for the $s=2,4$ cases fixing the leading terms by partial integration and field redefinition leads to the unique solution of Noether's equation (\ref{4.482}).

\section{Cubic selfinteraction and Noether's procedure, the spin two example}
\setcounter{equation}{0}
\quad
Using our general basis for the spin 2 case
\begin{eqnarray}
&&h_{\mu\nu},\\
&&D_{\mu}=(\nabla h)_{\mu}-\frac{1}{2}\nabla_{\mu}h \quad(\textnormal{de Donder term}),\\
&&h=h^{\ \mu}_{\mu} \quad (\textnormal{trace term}).
\end{eqnarray}
we can rewrite the free Fronsdal (linearized Einstein-Hilbert gravity) Lagrangian for the spin two gauge field in the following way:
\begin{eqnarray}
&&\mathcal{L}_{0}=-\frac{1}{2}h^{\mu\nu}(\Box h_{\mu\nu}
-2\nabla_{(\mu} D_{\nu)})+\frac{1}{4}h(\Box h-2(\nabla D)),\label{s2}\\
&& (\nabla D)=\nabla^{\mu}D_{\mu} .
\end{eqnarray}
This action is invariant with respect to the zero order gauge transformation
\begin{eqnarray}
\delta_{(0)}h_{\mu\nu}=2\nabla_{(\mu}\varepsilon_{\nu)}.
\end{eqnarray}

According to our strategy described in the previous section we obtain the following cubic interaction Lagrangian
\begin{eqnarray}
\mathcal{L}_{1}(h^{(2)})=&&\frac{1}{2}h^{\alpha\beta}\nabla_{\alpha}\nabla_{\beta}h_{\mu\nu}h^{\mu\nu}
                         +h^{\alpha\mu}\nabla_{\alpha}h^{\beta\nu}\nabla_{\beta}h_{\mu\nu}\nonumber\\
                         &&-\frac{1}{4}(\nabla D)h_{\mu\nu}h^{\mu\nu}
                         -\frac{1}{2}h^{\mu\nu}\nabla_{\mu}h D_{\nu},\label{intgravity}
\end{eqnarray}
supplemented with the Lie derivative form of the first order transformation law
\begin{eqnarray}
\delta_{(1)}h_{\mu\nu}=\varepsilon^{\rho}\nabla_{\rho}h_{\mu\nu}+2\nabla_{(\mu}\varepsilon^{\rho}h_{\nu)\rho}\label{deltagravity}
\end{eqnarray}
and the following field redefinition leading to this minimized form of Lagrangian (\ref{intgravity})
\begin{eqnarray}
h_{\mu\nu} \rightarrow h_{\mu\nu}+\frac{1}{4}(hh_{\mu\nu}-2h_{\mu}^{\ \rho}h_{\nu\rho}-\frac{1}{2(d-2)}h^{2}g_{\mu\nu})\label{frgravity}
\end{eqnarray}
Note that the interaction Lagrangian in de Donder gauge
\begin{eqnarray}
D_{\mu}=0,
\end{eqnarray}
reduces to the first two leading terms of (\ref{intgravity}). This minimized form of the leading terms is equivalent to the expansion up to cubic terms of the Einstein-Hilbert action (see formula (2.24) \cite{vanDam}) after partial integration and field redefinition, and is in full agreement with (\ref{prediction}) for $s=2$.

To see the same for the first order transformation law (\ref{deltagravity}) and (\ref{pred1}) we note that
the second term in the (\ref{deltagravity}) can be written in the form involving the vector curvature $\gamma^{(1)}_{\mu\nu}=2\nabla_{[\mu}\varepsilon_{\nu]}$ and the additional field redefinition
\begin{eqnarray}
(\nabla_{(\mu}\varepsilon^{\rho}-\nabla^{\rho}\varepsilon_{(\mu})h_{\nu)\rho}
+(\nabla_{(\mu}\varepsilon^{\rho}+\nabla^{\rho}\varepsilon_{(\mu})h_{\nu)\rho}\nonumber\\
=(\nabla_{(\mu}\varepsilon^{\rho}-\nabla^{\rho}\varepsilon_{(\mu})h_{\nu)\rho}
+\frac{1}{2}\delta_{\varepsilon}^{0}(h_{(\mu}^{\ \ \rho}h_{\nu)\rho})
\end{eqnarray}
Consequently the first order gauge variation becomes
\begin{eqnarray}
\delta_{(1)}h_{\mu\nu}=
\varepsilon^{\rho}\nabla_{\rho}h_{\mu\nu}
+\gamma^{(1)\,\rho}_{(\mu}h_{\nu)\rho},
\end{eqnarray}
and the field redefinition (\ref{frgravity}) reduces to
\begin{eqnarray}
h_{\mu\nu} \rightarrow h_{\mu\nu}+\frac{1}{4}(hh_{\mu\nu}-\frac{1}{2(d-2)}h^{2}g_{\mu\nu})\label{frspin2}
\end{eqnarray}

\section{The cubic selfinteraction for spin four}
\setcounter{equation}{0}
\quad We  start this nontrivial case by introducing the free Fronsdal's Lagrangian for the spin four gauge field $h_{\alpha\beta\gamma\delta}$
\begin{eqnarray}
&&\mathcal{L}_{0}(h^{(4)})=-\frac{1}{2}h^{\alpha\beta\gamma\delta}\mathcal{F}_{\alpha\beta\gamma\delta}
                         +\frac{3}{2}\bar{h}^{\alpha\beta}\bar{\mathcal{F}}_{\alpha\beta}\label{f41}\\
&&\mathcal{F}_{\alpha\beta\gamma\delta}=\Box h_{\alpha\beta\gamma\delta}-4\nabla_{(\alpha}D_{\beta\gamma\delta)}\label{f42}\\
&&\bar{\mathcal{F}}_{\alpha\beta}=\mathcal{F}^{\gamma}_{\gamma\alpha\beta}=\Box \bar{h}^{}_{\alpha\beta}-2(\nabla D)_{\alpha\beta}\label{f43}
\end{eqnarray}
which is invariant under
\begin{eqnarray}
\delta_{(0)}h_{\alpha\beta\gamma\delta}=4\nabla_{(\alpha}\epsilon_{\beta\gamma\delta)}
\end{eqnarray}
where we defined the de Donder tensor and the trace of the gauge field by
\begin{eqnarray}
&D_{\alpha\beta\gamma}=(\nabla h)_{\alpha\beta\gamma}-\frac{3}{2}\nabla_{(\alpha}\bar{h}^{}_{\beta\gamma)},\\
&\bar{h}^{}_{\beta\gamma}=h^{\ \ \ \ \alpha}_{\beta\gamma\alpha},\\
&D_{\alpha\beta}^{\ \ \beta}=0, \ \ \bar{h}^{\ \beta}_{\beta}=0.
\end{eqnarray}

The spin four case is much more complicated than the spin two case and includes all difficulties and complexities of a general spin $s$ interaction but remains inside the domain of problems which one can handle analytically. To apply our strategy and integrate the corresponding Noether's equation completely we have to introduce the following table to classify terms and levels of the interaction Lagrangian.
\begin{equation}\label{table}
    \setlength{\unitlength}{0.254mm}
\begin{picture}(351,360)(120,-440)
        %\special{color rgb 0 0 0}
        \allinethickness{0.254mm}\path(120,-80)(405,-80) % Plain Solid Line
        \allinethickness{0.254mm}\path(120,-80)(120,-440) % Plain Solid Line
        \allinethickness{0.254mm}\path(165,-80)(165,-440) % Plain Solid Line
        \allinethickness{0.254mm}\path(120,-120)(405,-120) % Plain Solid Line
        \allinethickness{0.254mm}\path(325,-80)(325,-360) % Plain Solid Line
        \allinethickness{0.254mm}\path(405,-80)(405,-280) % Plain Solid Line
        \allinethickness{0.254mm}\path(120,-200)(405,-200) % Plain Solid Line
        \allinethickness{0.254mm}\path(120,-280)(405,-280) % Plain Solid Line
        \allinethickness{0.254mm}\path(120,-360)(325,-360) % Plain Solid Line
        \allinethickness{0.254mm}\path(120,-440)(245,-440) % Plain Solid Line
        \allinethickness{0.254mm}\path(245,-80)(245,-440) % Plain Solid Line
        \allinethickness{0.254mm}\path(120,-80)(165,-120) % Plain Solid Line
        \put(200,-106){\shortstack{$0$}} % Plain Text
        \put(275,-106){\shortstack{$1$}} % Plain Text
        \put(355,-106){\shortstack{$2$}} % Plain Text
        \put(135,-166){\shortstack{$0$}} % Plain Text
        \put(135,-246){\shortstack{$1$}} % Plain Text
        \put(135,-331){\shortstack{$2$}} % Plain Text
        \put(135,-401){\shortstack{$3$}} % Plain Text
        \put(145,-96){\shortstack{$D$}} % Plain Text
        \put(135,-116){\shortstack{$\bar{h}$}} % Plain Text
        \put(190,-166){\shortstack{$hhh$}} % Plain Text
        \put(345,-166){\shortstack{$DDh$}} % Plain Text
        \put(270,-166){\shortstack{$Dhh$}} % Plain Text
        \put(190,-246){\shortstack{$\bar{h}hh$}} % Plain Text
        \put(190,-331){\shortstack{$\bar{h}\bar{h}h$}} % Plain Text
        \put(190,-401){\shortstack{$\bar{h}\bar{h}\bar{h}$}} % Plain Text
        \put(265,-246){\shortstack{$\bar{h}Dh$}} % Plain Text
        \put(340,-246){\shortstack{$\bar{h}DD$}} % Plain Text
        \put(265,-331){\shortstack{$\bar{h}\bar{h}D$}} % Plain Text
         % Set color to black again (default font color)
\end{picture}
\end{equation}
This table introduces some "coordinate system" for classification of our interaction
\begin{equation}\label{intlag}
    \mathcal{L}_{1}=\sum_{i,j=0,1,2,3 \atop  i+j\leq 3} \mathcal{L}^{\emph{int}}_{ij}(h^{(4)})
\end{equation}
where
\begin{equation}\label{ij}
    \mathcal{L}^{\emph{int}}_{ij}(h^{(4)}) \sim \nabla^{4-i} (D)^{i} (\bar{h}^{(4)})^{j} (h^{(4)})^{3-j-i}
\end{equation}
In this notation the leading term described in the second section is $\mathcal{L}^{\emph{int}}_{00}(h^{(4)})$. On the other hand the first column of table (\ref{table}) is nothing else but the interaction Lagrangian in de Donder gauge $D_{\alpha\beta\gamma}=0$ and can be expressed as a sum
\begin{eqnarray}
\mathcal{L}^{\emph{int}}_{dD}(h^{(4)})=\sum_{j=0}^{3} \mathcal{L}^{\emph{int}}_{0j}(h^{(4)})
\end{eqnarray}
Integrating Noether's equation step by step (cell by cell in means of (\ref{table})) starting from the initial curvature ansatz (\ref{impterm}), we obtain after very long and tedious calculations the following cubic interaction Lagrangian:
\begin{eqnarray}
\mathcal{L}^{\emph{int}}_{00}(h^{(4)})=
&&\ \frac{1}{8}h^{\alpha\beta\gamma\delta}h^{\mu\nu\lambda\rho}\Gamma_{\alpha\beta\gamma\delta,\mu\nu\lambda\rho}
-\nabla^{\mu}h_{\alpha\beta\gamma\delta}\nabla^{\alpha}\nabla^{\beta}h^{\gamma\nu\lambda\rho}\nabla^{\delta}h_{\mu\nu\lambda\rho}\nonumber\\
&&+\frac{3}{4}\nabla^{\mu}h_{\alpha\beta\gamma\delta}\nabla^{\alpha}\nabla^{\nu}h^{\gamma\delta\lambda\rho}\nabla^{\beta}h_{\mu\nu\lambda\rho}\nonumber\\
&&+3\nabla^{\mu}\nabla_{\nu}h_{\alpha\beta\gamma\delta}h^{\alpha\nu\lambda\rho}\nabla^{\beta}\nabla_{\lambda}h_{\mu\rho}^{\ \ \ \gamma\delta},\label{00}
\end{eqnarray}
\begin{eqnarray}
\mathcal{L}^{\emph{int}}_{01}(h^{(4)})=
&&-\frac{3}{2}h_{\alpha\beta\gamma\delta}\nabla^{\alpha}\nabla^{\beta}h^{\gamma\nu\lambda\rho}\nabla^{\delta}\nabla_{\nu}\bar{h}^{}_{\lambda\rho}
-3h_{\alpha\beta\gamma\delta}h_{\nu\lambda\rho}^{\ \ \ \delta}\nabla^{\alpha}\nabla^{\beta}\nabla^{\nu}\nabla^{\lambda}\bar{h}^{\gamma\rho}\nonumber\\
&&+\frac{3}{2}\nabla_{\mu}h_{\alpha\beta\gamma\delta}\nabla^{\nu}\nabla^{\alpha}h^{\mu\beta\gamma\lambda}\nabla^{\delta}\bar{h}^{}_{\nu\lambda}
-\nabla^{\lambda}h^{\mu\alpha\beta\gamma}\nabla^{\rho}h^{\nu}_{\ \alpha\beta\gamma}\nabla_{\mu}\nabla_{\nu}\bar{h}^{}_{\lambda\rho}\nonumber\\
&&+\frac{1}{4}h^{\mu\alpha\beta\gamma}h^{\nu}_{\ \alpha\beta\gamma}\nabla_{\mu}\nabla_{\nu}(\nabla\nabla \bar{h}^{}),
\end{eqnarray}
\begin{eqnarray}
\mathcal{L}^{\emph{int}}_{02}(h^{(4)})=
&&-\frac{3}{2}h_{\alpha\beta\gamma\delta}\nabla^{\alpha}\nabla^{\beta}\nabla^{\mu}\bar{h}^{\gamma\nu}\nabla^{\delta}\bar{h}^{}_{\mu\nu}
+\frac{3}{2}h_{\alpha\beta\gamma\delta}\nabla^{\alpha}\nabla^{\mu}\bar{h}^{\beta\nu}\nabla^{\gamma}\nabla_{\nu}\bar{h}^{\ \delta}_{\mu}\nonumber\\
&&-\frac{3}{4}\nabla_{\mu}\nabla_{\nu}h_{\alpha\beta\gamma\delta}\nabla^{\alpha}\bar{h}^{\beta\nu}\nabla^{\gamma}\bar{h}^{\delta\mu}
-\frac{3}{4}h_{\alpha\beta\gamma\delta}\nabla^{\alpha}\nabla^{\beta}\bar{h}^{\gamma\delta}(\nabla\nabla \bar{h}^{})\nonumber\\
&&-3\nabla_{\mu}h_{\alpha\beta\gamma\delta}\nabla_{\nu}\nabla^{\alpha}\bar{h}^{\beta\gamma}\nabla^{\delta}\bar{h}^{}_{\mu\nu},
\end{eqnarray}
\begin{eqnarray}
\mathcal{L}^{\emph{int}}_{03}(h^{(4)})=
\frac{3}{4}\nabla_{\mu}\nabla_{\nu}\bar{h}^{}_{\alpha\beta}\nabla^{\alpha}\bar{h}^{\mu\lambda}\nabla^{\beta}\bar{h}^{\nu}_{\ \lambda}
-\frac{3}{4}\nabla_{\mu}\bar{h}^{\nu\lambda}\nabla_{\nu}\bar{h}^{\mu}_{\ \lambda}(\nabla\nabla \bar{h}^{}),
\end{eqnarray}
\begin{eqnarray}
\mathcal{L}^{\emph{int}}_{10}(h^{(4)})=
&&\ \ 3\nabla_{\alpha}\nabla_{\nu}D_{\lambda\rho\beta}h^{\alpha\beta\gamma\delta}\nabla_{\gamma}h^{\ \nu\lambda\rho}_{\delta}
+\frac{3}{2}\nabla^{\rho}D_{\alpha\beta\lambda}\nabla^{\mu}h^{\alpha\beta\gamma\delta}\nabla^{\lambda}h_{\rho\mu\gamma\delta}\nonumber\\
&&-2\nabla^{\delta}D_{\nu\lambda\rho}\nabla^{\nu}h_{\alpha\beta\gamma\delta}\nabla^{\lambda}h^{\rho\alpha\beta\gamma}
-\frac{3}{2}(\nabla D)^{\alpha\rho}\nabla^{\mu}h_{\alpha\beta\gamma\delta}\nabla^{\beta}h_{\rho\mu}^{\ \ \gamma\delta}\nonumber\\
&&+\frac{1}{4}(\nabla D)^{\mu\nu}\nabla_{\mu}h_{\alpha\beta\gamma\delta}\nabla_{\nu}h^{\alpha\beta\gamma\delta}
-\frac{1}{2}(\nabla D)^{\mu\nu}h_{\alpha\beta\gamma\delta}\nabla_{\mu}\nabla_{\nu}h^{\alpha\beta\gamma\delta}\nonumber\\
&&-\nabla_{\alpha}(\nabla D)^{\mu\nu}h^{\alpha\beta\gamma\delta}\nabla_{\mu}h_{\nu\beta\gamma\delta}
+\frac{3}{4}\nabla_{\alpha}(\nabla D)^{\mu\nu}h^{\alpha\beta\gamma\delta}\nabla_{\beta}h_{\mu\nu\gamma\delta},
\end{eqnarray}
\begin{eqnarray}
\mathcal{L}^{\emph{int}}_{11}(h^{(4)})=
&&-\frac{1}{2}\bar{h}^{\gamma\delta}\nabla_{\gamma}h_{\mu\nu\lambda\rho}\nabla_{\delta}\nabla^{\mu}D^{\nu\lambda\rho}
+\frac{1}{2}\bar{h}^{\gamma\delta}\nabla_{\gamma}\nabla_{\delta}h_{\mu\nu\lambda\rho}\nabla^{\mu}D^{\nu\lambda\rho}\nonumber\\
&&+\frac{3}{4}\nabla^{\mu}\bar{h}^{\gamma\delta}h_{\mu\nu\lambda\rho}\nabla_{\gamma}\nabla^{\nu}D_{\delta}^{\ \lambda\rho}
-\frac{3}{4}\nabla_{\mu}\bar{h}^{\gamma\delta}h^{\mu\nu\lambda\rho}\nabla_{\nu}\nabla_{\lambda}D_{\gamma\delta\rho}\nonumber\\
&&+\frac{9}{4}\nabla_{\mu}\bar{h}^{\gamma\delta}\nabla_{\rho}h_{\gamma\delta\nu\lambda}\nabla^{\lambda}D^{\mu\nu\rho}
+3\bar{h}^{\gamma\delta}\nabla_{\rho}h_{\gamma\mu\nu\lambda}\nabla^{\mu}\nabla^{\nu}D_{\delta}^{\ \lambda\rho}\nonumber\\
&&+\frac{3}{2}\bar{h}^{\gamma\delta}\nabla_{\rho}h_{\gamma\mu\nu\lambda}\nabla^{\mu}\nabla_{\delta}D^{\nu\lambda\rho}
-\frac{3}{2}\bar{h}^{\gamma\delta}\nabla_{\rho}\nabla_{\gamma}h_{\delta\mu\nu\lambda}\nabla^{\mu}D^{\nu\lambda\rho}\nonumber\\
&&-\frac{3}{4}(\nabla D)^{\gamma\delta}\nabla^{\mu}\bar{h}^{\nu\lambda}\nabla_{\gamma}h_{\delta\mu\nu\lambda}
-\frac{3}{4}\nabla^{\mu}(\nabla D)^{\gamma\delta}\bar{h}^{\nu\lambda}\nabla_{\lambda}h_{\gamma\delta\mu\nu}\nonumber\\
&&+6\nabla^{\mu}\nabla_{\nu}(\nabla D)^{\gamma\delta}\bar{h}^{}_{\gamma\lambda}h_{\delta\mu}^{\ \ \nu\lambda}
+\frac{1}{4}\bar{h}^{\gamma\delta}\Box D^{\mu\nu\rho}\nabla_{\gamma}h_{\delta\mu\nu\rho}\nonumber\\
&&-\frac{3}{8}\bar{h}^{\gamma\delta}\Box D^{\mu\nu\rho}\nabla_{\mu}h_{\gamma\delta\nu\rho},
\end{eqnarray}
\begin{eqnarray}
\mathcal{L}^{\emph{int}}_{12}(h^{(4)})=
&&\ \ \frac{3}{4}D^{\mu\nu\rho}(\nabla \bar{h}^{})^{\delta}\nabla_{\delta}\nabla_{\mu}\bar{h}^{}_{\nu\rho}
-\frac{9}{8}\nabla_{\gamma}\nabla_{\delta}D^{\mu\nu\rho}\bar{h}^{\gamma\delta}\nabla_{\mu}\bar{h}^{}_{\nu\rho}\nonumber\\
&&-3D^{\mu\nu\rho}\bar{h}^{\gamma\delta}\nabla_{\gamma}\nabla_{\mu}\nabla_{\nu}\bar{h}^{}_{\delta\rho}
-3\nabla_{\mu}D_{\nu\rho\gamma}\nabla^{\nu}\bar{h}^{\gamma\delta}\nabla_{\delta}\bar{h}^{\mu\rho}\nonumber\\
&&-\frac{3}{2}(\nabla D)^{\gamma\delta}\nabla_{\gamma}\bar{h}^{\mu\nu}\nabla_{\delta}\bar{h}^{}_{\mu\nu}
-\frac{3}{8}(\nabla D)^{\gamma\delta}\bar{h}^{\mu\nu}\nabla_{\gamma}\nabla_{\delta}\bar{h}^{}_{\mu\nu}\nonumber\\
&&+\frac{3}{2}\nabla^{\mu}(\nabla D)^{\gamma\delta}\nabla_{\gamma}\bar{h}^{}_{\mu\nu}\bar{h}_{\delta}^{\ \nu}
-3(\nabla D)^{\gamma\delta}\nabla_{\gamma}\nabla_{\mu}\bar{h}^{}_{\delta\nu}\bar{h}^{\mu\nu}\nonumber\\
&&-\frac{9}{4}(\nabla D)^{\gamma\delta}\nabla^{\mu}\bar{h}^{}_{\gamma\nu}\nabla^{\nu}\bar{h}^{}_{\delta\mu}
+\frac{3}{2}\nabla^{\mu}(\nabla D)^{\gamma\delta}\nabla^{\nu}\bar{h}^{}_{\gamma\delta}\bar{h}^{}_{\mu\nu}\nonumber\\
&&+\frac{3}{8}(\nabla D)^{\gamma\delta}\bar{h}^{}_{\gamma\delta}(\nabla\nabla \bar{h}^{})
+\frac{3}{8}\nabla^{\mu}\nabla^{\nu}(\nabla D)^{\gamma\delta}\bar{h}^{}_{\gamma\mu}\bar{h}^{}_{\delta\nu}\nonumber\\
&&-\frac{3}{2}\Box(\nabla D)^{\gamma\delta}\bar{h}^{}_{\gamma\mu}\bar{h}^{\ \mu}_{\delta},
\end{eqnarray}
\begin{eqnarray}
\mathcal{L}^{\emph{int}}_{20}(h^{(4)})=
&&\ \ 3D^{\alpha\beta\gamma}D^{\mu\nu\rho}\nabla_{\mu}\nabla_{\nu}h_{\rho\alpha\beta\gamma}
-\frac{9}{4}D^{\alpha\beta\gamma}D^{\mu\nu\rho}\nabla_{\alpha}\nabla_{\mu}h_{\beta\gamma\nu\rho}\nonumber\\
&&+3D^{\alpha\beta\gamma}\nabla^{\rho}D_{\gamma}^{\ \mu\nu}\nabla_{\alpha}h_{\beta\mu\nu\rho}
+\frac{1}{2}(\nabla D)^{\gamma\delta}D^{\mu\nu\rho}\nabla_{\gamma}h_{\delta\mu\nu\rho},
\end{eqnarray}
\begin{eqnarray}
\mathcal{L}^{\emph{int}}_{21}(h^{(4)})=
&&-\frac{3}{4}\bar{h}^{\gamma\delta}\nabla_{\gamma}D^{\mu\nu\rho}\nabla_{\delta}D_{\mu\nu\rho}
+\frac{1}{8}(\nabla\nabla \bar{h}^{})D^{\mu\nu\rho}D_{\mu\nu\rho}\nonumber\\
&&+\frac{3}{4}(\nabla \bar{h}^{})^{\delta}D^{\mu\nu\rho}\nabla_{\mu}D_{\delta\nu\rho}
+\frac{9}{4}\bar{h}^{\gamma\delta}\nabla_{\gamma}D^{\mu\nu\rho}\nabla_{\mu}D_{\delta\nu\rho}\nonumber\\
&&+3\bar{h}^{\gamma\delta}\nabla^{\mu}D_{\gamma}^{\ \nu\rho}\nabla_{\nu}D_{\delta\mu\rho}
+3\nabla^{\mu}\nabla_{\nu}\bar{h}^{\gamma\delta}D_{\gamma}^{\ \nu\rho}D_{\delta\mu\rho}\nonumber\\
&&-\frac{3}{4}\bar{h}^{\gamma\delta}(\nabla D)^{\mu\nu}\nabla_{\mu}D_{\nu\gamma\delta}
+6\bar{h}^{\gamma\delta}\nabla^{\mu}(\nabla D)_{\gamma}^{\ \nu}D_{\delta\mu\nu}\nonumber\\
&&+3\bar{h}^{\gamma\delta}(\nabla D)_{\gamma}^{\ \nu}(\nabla D)_{\delta\nu}
\end{eqnarray}

Collecting factors coming with Fronsdal's equation of motion (Fronsdal's tensor) in Noether's equation we obtain
next to the free term $\delta_{(0)}h$ of the gauge transformation law for the spin four field the linear term
\begin{eqnarray}
\delta_{(1)}h_{\alpha\beta\gamma\delta}=
&&\epsilon^{\mu\nu\rho}\nabla_{\mu}\nabla_{\nu}\nabla_{\rho}h_{\alpha\beta\gamma\delta}\nonumber\\
&&+3(\nabla_{\alpha}\epsilon_{\rho}^{\ \mu\nu}-\nabla_{\rho}\epsilon_{\alpha}^{\ \mu\nu})
\nabla_{\mu}\nabla_{\nu}h_{\beta\gamma\delta}^{\ \ \ \ \rho}\nonumber\\
&&+3(\nabla_{\alpha}\nabla_{\beta}\epsilon_{\nu\rho}^{\ \ \ \mu}
-2\nabla_{\alpha}\nabla_{\nu}\epsilon_{\beta\rho}^{\ \ \ \mu}
+\nabla_{\nu}\nabla_{\rho}\epsilon_{\alpha\beta}^{\ \ \ \mu})\nabla_{\mu}h_{\gamma\delta}^{\ \ \ \nu\rho}\nonumber\\
&&+(\nabla_{\alpha}\nabla_{\beta}\nabla_{\gamma}\epsilon_{\mu\nu\rho}
-3\nabla_{\alpha}\nabla_{\beta}\nabla_{\mu}\epsilon_{\gamma\nu\rho}
+3\nabla_{\alpha}\nabla_{\mu}\nabla_{\nu}\epsilon_{\beta\gamma\rho}
-\nabla_{\mu}\nabla_{\nu}\nabla_{\rho}\epsilon_{\alpha\beta\gamma})h_{\delta}^{\ \ \mu\nu\rho}\nonumber\\
&&+(trace\  terms\  O(g_{\alpha\beta}))\nonumber\\
=&&\gamma^{(0)\mu\nu\rho}_{(\epsilon^{(3)})}\nabla_{\mu}\nabla_{\nu}\nabla_{\rho}h_{\alpha\beta\gamma\delta}
+3\gamma_{(\epsilon^{(3)})\alpha,\rho}^{(1)\ \ \ \ \ \mu\nu}\nabla_{\mu}\nabla_{\nu}h^{\ \ \ \ \rho}_{\beta\gamma\delta}\nonumber\\
&&+3\gamma_{(\epsilon^{(3)})\alpha\beta,\nu\rho}^{(2)\ \ \ \ \ \ \ \ \mu}\nabla_{\mu}h^{\ \ \ \nu\rho}_{\gamma\delta}
+\gamma_{(\epsilon^{(3)})\alpha\beta\gamma,\mu\nu\rho}^{(3)}h^{\ \mu\nu\rho}_{\delta}\nonumber\\
&&+(trace\  terms\  O(g_{\alpha\beta})),\label{translow}
\end{eqnarray}
where we assumed symmetrization of the indices $\alpha,\beta,\gamma,\delta$
and the spin four field redefinition
\begin{eqnarray}
h_{\alpha\beta\gamma\delta} \rightarrow h_{\alpha\beta\gamma\delta}
&&-\frac{9}{8}\nabla_{\mu}\nabla_{\nu}\bar{h}_{\alpha\beta}h_{\gamma\delta}^{\,\,\,\,\,\,\mu\nu}
-\frac{1}{4}(\nabla\nabla\bar{h})h_{\alpha\beta\gamma\delta}
-\frac{3}{4}\nabla_{\mu}\left[(\nabla\bar{h})^{\mu}h_{\alpha\beta\gamma\delta}\right]\nonumber\\
&&+\frac{1}{2}\bar{h}^{\mu\nu}\nabla_{\mu}\nabla_{\alpha}h_{\beta\gamma\delta\nu}
+\nabla_{\nu}(\nabla\bar{h})_{\alpha}h_{\beta\gamma\delta}^{\,\,\,\,\,\,\,\,\,\,\nu}
-\frac{3}{2}\nabla_{\mu}\bar{h}_{\nu\alpha}\nabla_{\beta}h_{\gamma\delta}^{\,\,\,\,\,\,\mu\nu}\nonumber\\
&&-\frac{3}{8}\bar{h}^{\mu\nu}\nabla_{\alpha}\nabla_{\beta}h_{\gamma\delta\mu\nu}
+\frac{1}{4}\nabla^{\mu}(\bar{h}_{\mu\alpha}D_{\beta\gamma\delta}-\frac{3}{2}\bar{h}_{\alpha\beta}D_{\gamma\delta\mu})\nonumber\\
&&+\frac{9}{2}\nabla^{\mu}\nabla_{\alpha}\bar{h}_{\beta\gamma}\bar{h}_{\delta\mu}
-\frac{21}{32}\nabla_{\nu}\bar{h}_{\alpha\beta}\nabla^{\nu}\bar{h}_{\gamma\delta}\nonumber\\
&&-\frac{3}{2}\nabla_{\alpha}\nabla_{\beta}\bar{h}_{\gamma}^{\,\,\,\mu}\bar{h}_{\delta\mu}
+\frac{15}{8}(\nabla\bar{h})_{\alpha}\nabla_{\beta}\bar{h}_{\gamma\delta}\nonumber\\
&&+(trace\  terms\  O(g_{\alpha\beta})),
\end{eqnarray}
where symmetrization over the indices $\alpha,\beta,\gamma,\delta$ is also understood.

Finally note that we did not obtain an $\mathcal{L}^{\emph{inter}}_{30} \sim (D)^{3}$ part of interaction (there is no corresponding cell in the first row of (\ref{table})) because we started the leading part $\mathcal{L}^{\emph{inter}}_{00}$ (\ref{00}) from the curvature term and fixed in this way partial integration freedom. After that as it was mentioned above all other terms of interaction could be constructed in a unique way up to some field redefinition. This particular way of derivative rearrangement (including partial integration of all other level terms)  does not lead to a $(D)^{3}$ term as opposed to other ways of rearranging the derivatives by means of the partial integration freedom.
On the other hand if we rearrange the derivatives as described in the second section we get as leading part of the interaction $\mathcal{L}^{\emph{inter}}_{00}$ in complete agreement with our prediction (\ref{prediction}) for $s=4$. The same is true for the transformation law (\ref{translow}) and (\ref{pred1}).

\section{Conclusion}
Based on an algorithmic and partially recursive construction scheme both the cubic selfinteraction of the spin four higher spin field
and that part of the gauge transformation that is linear in the gauge field (first order gauge transformation) were derived. For
general even spin a formula for the leading part (containing no traces, divergences or de Donder terms) of the cubic interaction
and the corresponding form of the first order gauge transformation were guessed. The similarity of these results with gravity theory
was worked out. Obviously these results open the door to many investigations, in particular also to an extended gravity theory.

\subsection*{Acknowledgements}
This work is supported in part by Alexander von Humboldt Foundation under 3.4-Fokoop-ARM/1059429 and ANSEF 2009.
Work of K.M. was made with partial support of CRDF-NFSAT UCEP06/07.

\section*{Appendix A\\ Higher spin free fields in Fronsdal's formulation and the deWit-Friedman linearized curvatures}
\setcounter{equation}{0}
\renewcommand{\theequation}{A.\arabic{equation}}
\quad
We will use the deWit-Freedman curvature and Cristoffel symbols \cite{DF, MR6}. We contract them with the degree $s$
tensorial power of one tangential vector $a^{\mu}$ in the first set of $s$ indices  and with a similar tensorial power of another tangential vector $b^{\nu}$ in its second set. The deWit-Freedman curvature and n-th Cristoffel symbol  are then written as
\begin{eqnarray}
&&\Gamma^{(s)}(z;b,a):\qquad\Gamma^{(s)}(z; b,\lambda a) = \Gamma^{(s)}(z;\lambda b, a)= \lambda^{s}\Gamma^{(s)}(z;b, a) ,\\
&&\Gamma^{(s)}_{(n)}(z;b,a):\qquad\Gamma^{(s)}_{(n)}(z;b,\lambda a) = \lambda^{s}\Gamma^{(s)}_{(n)}(z;b, a) ,\\
&&\quad\quad\quad\quad\quad\quad\quad\,\,\,\,\Gamma^{(s)}_{(n)}(z; \lambda b, a)=\lambda^{n}\Gamma^{(s)}_{(n)}(z;b, a) ,\\
&&\Gamma^{(s)}(z;b,a)=\Gamma^{(s)}_{(n)}(z;b,a)|_{n=s} .
\label{2.6}
\end{eqnarray}
Next we introduce the notation $*_a, *_b$ for a contraction in the symmetric spaces of indices $a$ or $b$
\begin{eqnarray}
  *_{a}&=&\frac{1}{(s!)^{2}} \prod^{s}_{i=1}\overleftarrow{\partial}^{\mu_{i}}_{a}\overrightarrow{\partial}_{\mu_{i}}^{a} .
   \label{4.12}
\end{eqnarray}

To manipulate reshuffling of different sets of indices we employ two differentials with respect to $a$ and $b$, e.g.
\begin{eqnarray}
A_{b} = (a\partial_{b}) , \\
\label{2.11}
B_{a} = (b\partial_{a}) .
\label{2.12}
\end{eqnarray}
Then we see that the operators $A_{b}, a^{2}, b^{2}$ are dual (or adjoint) to $B_{a},\Box_{a},\Box_{b}$ with respect to the "star" product of tensors with two sets of symmetrized indices  (\ref{4.12})
\begin{eqnarray}
    \frac{1}{n}A_{b}f^{(m-1,n)}(a,b)*_{a,b} g^{(m,n-1)}(a,b)&=& f^{(m-1,n)}(a,b)*_{a,b} \frac{1}{m}B_{a}g^{(m,n-1)}(a,b) ,\label{4.13}\\
    a^{2}f^{(m-2,n)}(a,b)*_{a,b} g^{(m,n)}(a,b)&=&f^{(m-2,n)}(a,b)*_{a,b} \frac{1}{m(m-1)}\Box_{a} g^{(m,n)}(a,b) . \nonumber\\\label{4.14}
\end{eqnarray}
In the same fashion gradients and divergences are dual with respect to the full scalar product in the space $(z,a,b)$
\begin{eqnarray}
  (a\nabla)f^{(m-1,n)}(z;a,b)*_{a,b} g^{(m,n)}(z;a,b) &=& -f^{(m-1,n)}(z;a,b)*_{a,b}\frac{1}{m}(\nabla\partial_{a}) g^{(m,n)}(z;a,b) .\nonumber\\ \label{4.15}
  \end{eqnarray}
Analogous equations can be formulated for the operators $b^{2}$ or $b\nabla$.

Now one can prove that \cite{DF, MR5}:
\begin{equation}
A_{b}\Gamma^{(s)}(z;a,b) = B_{a}\Gamma^{(s)}(z;a,b) = 0 .
\label{4.18}
\end{equation}
These "primary Bianchi identities" are manifestations of the hidden antisymmetry.
The n-th deWit-Freedman-Cristoffel symbol is
\begin{eqnarray}
\Gamma_{(n)}^{(s)}(z;b,a)&&\equiv \Gamma^{(s)}_{(n)\rho_{1}...\rho_{n},\mu_{1}...\mu_{\ell}}b^{\rho_{1}}...b^{\rho_{n}}
a^{\mu_{1}}...a^{\mu_{\ell}}\nonumber\\&&=[(b\nabla)-\frac{1}{n}(a\nabla)B_{a}]\Gamma_{(n-1)}^{(s)}(z;b,a) ,
\end{eqnarray}
or in another way
\begin{equation}
\Gamma_{(n)}^{(s)}(z;b,a)=(\prod_{k=1}^{s}[(b\nabla)-\frac{1}{k}(a\nabla)B_{a}])h^{(s)}(z;a) .
\end{equation}
Using the following commutation relations
\begin{eqnarray}
&&[B_{a},(a\nabla)]=(b\nabla),\label{4.21}\\
&&[B_{a}^{k},(a\nabla)]=kB_{a}^{k-1}(b\nabla),\\
&&[B_{a},(a\nabla)^{k}]=k(b\nabla)(a\nabla)^{k-1},\\
&&\Box_{b}(b\nabla)^{i}=i(i-1)(b\nabla)^{i-2}\Box,\\
&&\partial^{b}_{\mu}(b\nabla)^{i}\partial_{b}^{\mu}B_{a}^{j}=ij(b\nabla)^{i-1}B_{a}^{j-1}(\nabla\partial_{a}),\\
&&\Box_{b}B_{a}^{j}=j(j-1)B_{a}^{j-2}\Box_{a} ,\label{4.26}
\end{eqnarray}
and mathematical induction we can prove that
\begin{eqnarray}
\Gamma_{(n)}^{(s)}(z;b,a)=\sum_{k=0}^{n}\frac{(-1)^{k}}{k!}(b\nabla)^{n-k}(a\nabla)^{k}B_{a}^{k}h^{(s)}(z;a) .\label{gumar}
\end{eqnarray}
The gauge variation of the n-th Cristoffel symbol is
\begin{eqnarray}
&&\delta \Gamma_{(n)}^{(s)}(z;b,a)=\frac{(-1)^{n}}{n!}(a\nabla)^{n+1}B_{a}^{n}\epsilon^{(s-1)}(z;a) ,
\end{eqnarray}
putting here $n=s$ we obtain gauge invariance for the curvature
\begin{equation}\label{A22}
\delta \Gamma_{(s)}^{(s)}(z;b,a)=0 .
\end{equation}

Tracelessness of the gauge parameter (\ref{4.6})
implies that b-traces of all Cristoffel symbols are gauge invariant
\begin{eqnarray}
&& \Box_{b}\delta \Gamma_{(n)}^{(s)}(z;b,a)=\frac{(-1)^{n}}{(n-2)!}(a\nabla)^{n+1}B_{a}^{n-2}\Box_{a}\epsilon^{(s-1)}(z;a)=0 .
\end{eqnarray}
Thus for the second order gauge invariant field equation we can use the trace of the second Cristoffel symbol,
the so-called Fronsdal tensor:
\begin{eqnarray}
\mathcal{F}^{(s)}(z;a)&=&\frac{1}{2}\Box_{b}\Gamma_{(2)}^{(s)}(z;b,a)\nonumber\\
&=&\Box h^{(s)}(z;a)-(a\nabla)(\nabla\partial_{a})h^{(s)}(z;a)
+\frac{1}{2}(a\nabla)^{2}\Box_{a}h^{(s)}(z;a) .\quad\label{4.32}
\end{eqnarray}
Using equation (\ref{gumar}) for Cristoffel symbols and
after long calculations we obtain the following expression
\begin{eqnarray}
&&\Box_{b}\Gamma_{(n)}^{(s)}(z;b,a)\nonumber\\&&=\sum_{k=0}^{n-2}\frac{(-1)^{k}}{k!}(n-k)(n-k-1)
(b\nabla)^{n-k-2}(a\nabla)^{k}B_{a}^{k}\mathcal{F}^{(s)}(z;a) .
\end{eqnarray}
In particular for the trace of the curvature we can write
\begin{eqnarray}
&&\Box_{b}\Gamma^{(s)}(z;b,a)=s(s-1)\mathcal{U}(a,b,3,s)\mathcal{F}^{(s)}(z;a) ,\label{4.35}
\end{eqnarray}
where we introduced an operator mapping the Fronsdal tensor on the trace of the curvature
\begin{equation}\label{4.36}
   \mathcal{U}(a,b,3,s)=\prod^{s}_{k=3}[(b\nabla)-\frac{1}{k}(a\nabla)B_{a}] .
\end{equation}
Now let us consider this curvature in more detail. First we have the symmetry under exchange of $a$ and $b$
\begin{equation}
\Gamma^{(s)}(z;a,b) = \Gamma^{(s)}(z;b,a) .
\end{equation}
Therefore the operation "$a$-trace" can be defined by (\ref{4.35}) with exchange of $a$ and $b$ at the end.
The mixed trace of the curvature can be expressed through the $a$ or $b$ traces using "primary Bianchi identities" (\ref{4.18})
\begin{equation}\label{4.38}
(\partial_{a}\partial_{b})\Gamma^{(s)}(z;b,a)=-\frac{1}{2}B_{a}\Box_{b}\Gamma^{(s)}(z;b,a)=
-\frac{1}{2}A_{b}\Box_{a}\Gamma^{(s)}(z;b,a) .
\end{equation}

The next interesting properties of the higher spin curvature and corresponding Ricci tensors are so called generalized secondary or differential Bianchi identities. We can formulate  these identities  in our notation in the following compressed form ($[\dots]$ denotes antisymmetrization )
\begin{equation}\label{4.39}
\frac{\partial}{\partial a^{[\mu}}\frac{\partial}{\partial b^{\nu}}\nabla_{\lambda]}\Gamma^{(s)}(z;a,b)= 0 .
\end{equation}
This relation can be checked directly from representation (\ref{gumar}). Then contracting with $a^{\mu}$ and $b^{\nu}$ we get a symmetrized form of (\ref{4.39})
\begin{equation}\label{4.40}
    s \nabla_{\mu}\Gamma^{(s)}(z;a,b)=(a\nabla)\partial^{a}_{\mu}\Gamma^{(s)}(z;a,b)+(b\nabla)\partial^{b}_{\mu}\Gamma^{(s)}(z;a,b) .
\end{equation}
Now we can contract (\ref{4.40}) with a $\partial^{\mu}_{b}$ and using (\ref{4.38}) obtain a connection between the divergence and the trace of the curvature
\begin{equation}\label{4.41}
    (s-1)(\nabla\partial_{b})\Gamma^{(s)}(z;a,b)=[(b\nabla)-\frac{1}{2}(a\nabla)B_{a}]\Box_{b}\Gamma^{(s)}(z;a,b) .
\end{equation}
These two identities with a similar identity for the Fronsdal tensor
\begin{equation}\label{4.42}
    (\nabla\partial_{a})\mathcal{F}^{(s)}(z;a)=\frac{1}{2}(a\nabla)\Box_{a}\mathcal{F}^{(s)}(z;a) ,
\end{equation}
play an important role for the construction of the interaction Lagrangian.


\begin{thebibliography}{100}
\bibitem{M}
  R.~Manvelyan, K.~Mkrtchyan and W.~R\"uhl,
  ``Off-shell construction of some trilinear higher spin gauge field
  interactions,''
  Nucl.\ Phys.\  B {\bf 826} (2010) 1
  [arXiv:0903.0243 [hep-th]].
\bibitem{MMR}
R.~Manvelyan and K.~Mkrtchyan,
``Conformal invariant interaction of a scalar field with the higher spin
  field in $AdS_{D}$,''Mod.\ Phys.\ Lett.\  A {\bf 25} (2010) 1333, [arXiv:0903.0058 [hep-th]].
\bibitem{MR}
R.~Manvelyan and W.~R\"uhl, ``Conformal coupling of higher spin
gauge fields to a scalar field in AdS(4) and generalized Weyl
invariance,'' Phys.\ Lett.\ B {\bf 593} (2004) 253,
[arXiv:hep-th/0403241].
\bibitem{vanDam}
  F.~A.~Berends, G.~J.~H.~Burgers and H.~van Dam,
  ``Explicit Construction Of Conserved Currents For Massless Fields Of
  Arbitrary Spin,'' Nucl.\ Phys.\  B {\bf 271} (1986) 429;
   F.~A.~Berends, G.~J.~H.~Burgers and H.~Van Dam,
  ``On Spin Three Selfinteractions,''
  Z.\ Phys.\  C {\bf 24} (1984) 247;
  F.~A.~Berends, G.~J.~H.~Burgers and H.~van Dam,
  ``On The Theoretical Problems In Constructing Interactions Involving Higher
  Spin Massless Particles,''
  Nucl.\ Phys.\  B {\bf 260} (1985) 295.
\bibitem{Vasiliev}
  E.~S.~Fradkin and M.~A.~Vasiliev, ``On The Gravitational Interaction
  Of Massless Higher Spin Fields,'' Phys.\ Lett.\ B {\bf 189} (1987)
  89.
   E.~S.~Fradkin and M.~A.~Vasiliev, ``Cubic Interaction In
  Extended Theories Of Massless Higher Spin Fields,'' Nucl.\ Phys.\ B
  {\bf 291} (1987) 141.
\bibitem{Damour}
  T.~Damour and S.~Deser,
  ``Geometry of spin 3 gauge theories,''
  Annales Poincare Phys.\ Theor.\  {\bf 47}, 277 (1987);
 T.~Damour and S.~Deser,
  ``Higher derivative interactions of higher spin gauge fields,''
  Class.\ Quant.\ Grav.\  {\bf 4}, L95 (1987).
\bibitem{review}
M.A.~Vasiliev, ``Higher Spin Gauge Theories in Various Dimensions'',
Fortsch. Phys. 52, 702 (2004) [arXiv:hep-th/0401177].
 X.~Bekaert, S.~Cnockaert, C.~Iazeolla and M.A.~Vasiliev,
 ``Nonlinear higher spin theories in various dimensions
'', [arXiv:hep-th/0503128].
 D.~Sorokin,``Introduction to the Classical Theory of Higher Spins'' AIP Conf. Proc. 767, 172
(2005); [arXiv:hep-th/0405069]. N.~Bouatta, G.~Compere and A.~Sagnotti, ``An Introduction to Free Higher-Spin Fields''; [arXiv:hep-th/0409068].
\bibitem{Metsaev}
  R.~R.~Metsaev,
  ``Cubic interaction vertices for massive and massless higher spin fields,''
  Nucl.\ Phys.\  B {\bf 759} (2006) 147
  [arXiv:hep-th/0512342];R.~R.~Metsaev,
  ``Cubic interaction vertices for fermionic and bosonic arbitrary spin
  fields,''
  arXiv:0712.3526 [hep-th].
\bibitem{ouvry}
  I.~G.~Koh, S.~Ouvry, ``Interacting gauge fields of any spin and symmetry,''
  Phys. Lett. B {\bf 179} (1986) 115; Erratum-ibid. {\bf 183} B (1987) 434.
\bibitem{boulanger}
  N.~Boulanger, S.~Leclercq, P.~Sundell,
  ``On The Uniqueness of Minimal Coupling in Higher-Spin Gauge Theory,''
  JHEP 0808:056,2008;  [arXiv:0805.2764 [hep-th]].
  X.~Bekaert, N.~Boulanger, S.~Cnockaert, S.~Leclercq,
  ``On Killing tensors and cubic vertices in higher-spin gauge theories,''
  Fortsch. Phys. {\bf 54} (2006) 282-290; [arXiv:hep-th/0602092].
\bibitem{bekaert}
  X.~Bekaert, N.~Boulanger, S.~Leclercq, ``Strong obstruction of the Berends-Burgers-vanDam spin-3 vertex'',
  [arXiv:1002.0289[hep-th]]
\bibitem{Petkou}
  A.~Fotopoulos, N.~Irges, A.~C.~Petkou and M.~Tsulaia,
  ``Higher-Spin Gauge Fields Interacting with Scalars: The Lagrangian Cubic
  Vertex,''
  JHEP {\bf 0710} (2007) 021;
  [arXiv:0708.1399 [hep-th]].
  I.~L.~Buchbinder, A.~Fotopoulos, A.~C.~Petkou and M.~Tsulaia,
  ``Constructing the cubic interaction vertex of higher spin gauge fields,''
  Phys.\ Rev.\  D {\bf 74} (2006) 105018;
  [arXiv:hep-th/0609082].
 \bibitem{Deser}
  S.~Deser and Z.~Yang,
  ``Inconsistency of spin 4 - spin-2 gauge field couplings,''
  Class.\ Quant.\ Grav.\  {\bf 7} (1990) 1491.
\bibitem{ourlast}
  R.~Manvelyan, K.~Mkrtchyan and W.~R\"uhl,
    "General trilinear interaction for arbitrary even higher spin gauge fields",
      Nucl.\ Phys.\  B {\bf 836} (2010) 204, arXiv:1003.2877 [hep-th].
\bibitem{Kleb}
  I.~R.~Klebanov and A.~M.~Polyakov, ``AdS dual of the critical O(N)
  vector model,'' Phys.\ Lett.\ B {\bf 550} (2002) 213;
  [arXiv:hep-th/0210114].
\bibitem{MMR1}
R.~Manvelyan, K.~Mkrtchyan and W.~R\"uhl,
  ``Ultraviolet behaviour of higher spin gauge field propagators and one loop
  mass renormalization,''
  Nucl.\ Phys.\  B {\bf 803} (2008) 405
  [arXiv:0804.1211 [hep-th]].
\bibitem{Ruehl}
  W.~R\"uhl, ``The masses of gauge fields in higher spin field theory on
  AdS(4),'' Phys.Lett. B {\bf 605} (2005) 413; [arXiv:hep-th/0409252]; the results
  presented here are based on extensive calculations performed by K. Lang and
  W.~R\"uhl, Nucl. Phys. B {\bf 400} (1993) 597.
\bibitem{MR1}
  R.~Manvelyan and W.~R\"uhl,
  ``The off-shell behaviour of propagators and the Goldstone field in  higher
  spin gauge theory on AdS(d+1) space,''
  Nucl.\ Phys.\  B {\bf 717} (2005) 3;
  [arXiv:hep-th/0502123].
 \bibitem{MR2}
 R.~Manvelyan and W.~R\"uhl,
  ``The masses of gauge fields in higher spin field theory on the bulk of
  AdS(4),''
  Phys.\ Lett.\  B {\bf 613} (2005) 197;
  [arXiv:hep-th/0412252].
 \bibitem{MR3}
  R.~Manvelyan and W.~R\"uhl,
  ``The structure of the trace anomaly of higher spin conformal currents in the
  bulk of AdS(4),''
  Nucl.\ Phys.\  B {\bf 751}, (2006) 285;
  [arXiv:hep-th/0602067].
\bibitem{MR4} R.~Manvelyan and W.~R\"uhl,
  ``The quantum one loop trace anomaly of the higher spin conformal  conserved
  currents in the bulk of AdS(4),''
  Nucl.\ Phys.\  B {\bf 733} (2006) 104;
  [arXiv:hep-th/0506185].
\bibitem{MR5}
  R.~Manvelyan and W.~R\"uhl,
  ``Generalized Curvature and Ricci Tensors for a Higher Spin Potential and the
  Trace Anomaly in External Higher Spin Fields in $AdS_{4}$ Space,''
  Nucl.\ Phys.\  B {\bf 796} (2008) 457;
  [arXiv:0710.0952 [hep-th]].
\bibitem{Frons}
C.~Fronsdal,
 ``Singletons And Massless, Integral Spin Fields On De Sitter Space (Elementary
Particles In A Curved Space Vii),'' Phys.\ Rev.\ D {\bf 20},
(1979) 848;``Massless Fields With Integer Spin,'' Phys.\ Rev.\ D {\bf
18} (1978) 3624.
\bibitem{DF}
B.~deWit and D.Z.~Freedman, ``Systematics of higher spin gauge
fields,'' Phys. Review D \textbf{21} (1980), 358-367.
\bibitem{MR6}
  R.~Manvelyan and W.~R\"uhl,
  ``The Generalized Curvature and Christoffel Symbols for a Higher Spin
  Potential in $AdS_{d+1}$ Space,''
  Nucl.\ Phys.\  B {\bf 797}, 371 (2008)
  [arXiv:0705.3528 [hep-th]].
\bibitem{Giombi}
  S.~Giombi, Xi Yin, ``Higher spin gauge theory and holography: The three-point functions''
  [arXiv:0912.3462v3 [hep-th]].
  %\cite{Giombi:2010vg}
%\bibitem{Giombi:2010vg}
  S.~Giombi and X.~Yin,
  ``Higher Spins in AdS and Twistorial Holography,''
  arXiv:1004.3736 [hep-th].
  %%CITATION = ARXIV:1004.3736;%%



\end{thebibliography}
\end{document}